\documentclass[aps,prd,twocolumn,superscriptaddress,showpacs,showkeys,
nofootinbib]{revtex4-1}
\usepackage{latexsym}
\usepackage{amsmath}
\usepackage{amssymb}
\usepackage{amsfonts}
\usepackage{hyperref}
\usepackage{color}

\usepackage{supertabular} 
\usepackage{placeins}
\usepackage{epsfig}
\usepackage{graphicx}

\usepackage{epstopdf}

\begin{document}

%\preprint{}

\title{Charmonium resonances in the 3.9 GeV/$c^2$ energy region
and the $X(3915)/X(3930)$ puzzle}



\author{Pablo G. Ortega}
\email[]{pgortega@usal.es}
%\thanks{}
\affiliation{Grupo de F\'isica Nuclear and Instituto Universitario de F\'isica 
Fundamental y Matem\'aticas (IUFFyM), Universidad de Salamanca, E-37008 
Salamanca, Spain}

\author{Jorge Segovia}
\email[]{jorge.segovia@tum.de}
%\thanks{}
\affiliation{Physik-Department, Technische Universit\"at M\"unchen, 
James-Franck-Str.~1, 85748 Garching, Germany}

\author{David R. Entem}
\email[]{entem@usal.es}
%\thanks{}
%\affiliation{}

\author{Francisco Fern\'andez}
\email[]{fdz@usal.es}
%\thanks{}
\affiliation{Grupo de F\'isica Nuclear and Instituto Universitario de F\'isica 
Fundamental y Matem\'aticas (IUFFyM), Universidad de Salamanca, E-37008 
Salamanca, Spain}


\date{\today}

\begin{abstract}

An interesting controversy has emerged challenging the widely accepted nature of the $X(3915)$ 
and the $X(3930)$ resonances, which had initially been assigned to the $\chi_{c0}(2P)$
and $\chi_{c2}(2P)$ $c\bar c$ states, respectively.
To unveil their inner structure, the properties of the $J^{PC}\!\!=\!0^{++}$ and $J^{PC}\!\!=\!2^{++}$ charmonium states in the energy region of these resonances
are analyzed in the framework of a constituent quark model. Together with the bare $q\bar q$ states, threshold 
effects due to the opening of nearby meson-meson channels are included in a coupled-channels scheme calculation. 
We find that the structure of both states is dominantly molecular with a probability of bare $q\bar q$ states lower than $45\%$.
Our results favor the hypothesis that $X(3915)$ and $X(3930)$ resonances arise as different decay mechanisms of the same $J^{PC}\!\!=\!2^{++}$ state. 
Moreover we find an explanation for the recently discovered $M=3860$ MeV$/c^2$ as a $J^{PC}\!\!=\!0^{++}$ $2P$ state and rediscover the lost $Y(3940)$ as an additional state in the $J^{PC}\!\!=\!0^{++}$ family.
\end{abstract}

% insert suggested PACS numbers in braces on next line \phi


\maketitle

%%%%%%%%%%%%%%%%%%%%%%%%%%%%%%%%%%%%%%%%%%%%%%%%%%%%%%%%%%%%%%%%%%%%%%%%%%%%%%%
The region of the charmonium spectrum around 3.9 GeV/$c^2$, which would correspond with the  $\chi_{cJ}(2P)$ charmonium multiplet, is a very interesting one  
due to the presence of several unexpected states that do not fit into the predictions
of quark models.

The most famous state is the $X(3872)$ discovered in 2003 by the Belle Collaboration in the exclusive $B^\pm\to K^\pm \pi^+ \pi^- J/\psi$ decay ~\cite{Choi:2003ue}. 
This state decays through the $J/\psi\rho$ and $J/\psi\omega$ channels which are, respectively, forbidden and OZI-suppressed for a $c\bar c$ configuration.
Two years later a new state, called at that time $Y(3940)$, with a mass of $M=3943\pm 11\pm 13$ MeV$/c^2$ and a width of $\Gamma= 87\pm 22$ MeV was also reported by Belle in the decay $B^+\rightarrow K^+\omega J/\psi$~\cite{PhysRevLett.94.182002}. Additionally, in 2006, the same Collaboration found a peak in the mass spectrum of the $D\bar D$ mesons produced by $\gamma \gamma$ fusion. The values of mass and width of this state, originally named $Z(3930)$ and then $X(3930)$, were respectively
$M=3929\pm 6$ MeV$/c^2$ and $\Gamma=29\pm 10$ MeV. Finally, analyzing the double charmonium production in the reaction $e^+e^-\rightarrow J/\psi+X$, together with well-known charmonium states like the $\eta_c$, the $\chi_{c0}$ and the $\eta_c(2S)$, a new resonance, the $X(3940)$, with a mass of $M=3943\pm 8$ MeV$/c^2$ and a width of $\Gamma<52$ MeV was reported also by the Belle Collaboration~\cite{PhysRevLett.98.082001}.

The LHCb Experiment conclusively determined the $J^{PC}$ of the $X(3872)$ to be $1^{++}$ using a five-dimensional angular analysis of the process $B^+\rightarrow K^+X(3872)$ with $X(3872)\rightarrow J/\psi \rho^0\rightarrow J/\psi \pi^+\pi^-$~\cite{PhysRevD.92.011102}. The angular distribution of the $X(3930)$ in the $\gamma\gamma$ center of mass measured by Belle follows the one expected  for a $J=2$ state. Hence, the $X(3930)$ was rapidly assigned to the $\chi_{c2}(2P)$ charmonium state and incorporated to the PDG~\cite{Olive:2016xmw}, 
despite most of the quark models predict a mass higher than
the experimental one. For instance, the widely used Godfrey-Isgur relativistic quark model~\cite{Godfrey:1985xj} finds the aforementioned state at $M\!\!=\!\!3979$~MeV/$c^2$. 

The situation is worse in the case of the $X(3940)$ resonance. It has not been seen in the $D\bar D$ channel which rules out the $J^{PC}=0^{++}$ assignment. The dominance of the $\bar D D^*$ decay mode suggests that the $X(3940)$ is the $c\bar c (2^3P_1)$ state with $J^{PC}=1^{++}$, but these quantum numbers coincide with the ones of the $X(3872)$. In addition, a decay to $\omega J/\psi$ was not observed indicating that the $X(3940)$ and the $Y(3940)$ are not the same state. 
The history of the $Y(3940)$ is more complicated. In 2008, three years after its discovery, the Babar Collaboration claimed the confirmation of the $Y(3940)$ in the $B\rightarrow J/\psi \omega K$ decay, but with a mass somewhat smaller 
($3914$ MeV/$c^2$)~\cite{PhysRevLett.101.082001}. In 2010, the Belle Collaboration reported a resonance-like enhancement in the $\gamma\gamma\rightarrow\omega J/\psi$ process~\cite{Uehara:2009tx}, at $M=3915\pm 3\pm 2$ MeV/$c^2$ and $\Gamma=17\pm 10\pm 3$ MeV with possible quantum numbers $J^{PC}=0^{++}$ and $J^{PC}=2^{++}$. Finally, the BaBar Collaboration confirmed the existence of the $X(3915)$ and its spin-parity analysis clearly prefers the assignment $J^{PC}=0^{++}$~\cite{Lees:2012xs}. These authors pointed out that these values are consistent with those of the $Y(3940)$ and both signals are renamed as $X(3915)$. Then the state was eventually labeled as the $\chi_{c0}(2P)$ state by the 
PDG~\cite{Olive:2016xmw}. This assignment was also supported by the $\chi_{c0}(2P)$ mass value, $3916\,{\rm MeV}$, predicted by the 
Godfrey-Isgur relativistic quark model~\cite{Godfrey:1985xj}.

However problems do not end here. The $J^{PC}=0^{++}$ assignment was challenged by Guo and Meissner~\cite{Guo:2012tv} and also by Olsen~\cite{Olsen:2014maa}
mainly for three reasons:
\begin{itemize}
\item The partial width for the $X(3915)\rightarrow\omega J/\psi$ is too large for an OZI-suppressed decay.
\item There is not signal for the $X(3915)\rightarrow D\bar D$ decay, which is expected to be the dominant decay mechanism.
\item Assuming that the $X(3930)$ is the $\chi_{c2}(2P)$ state, the $\chi_{c2}(2P)-\chi_{c0}(2P)$ mass splitting is too small.
\end{itemize}

Beyond the discussion above, very recent studies have altered the previous situation. On the one hand, from the theoretical side, Z.-Y. Zhou \emph{et al.}~\cite{Zhou:2015uva}
revealed that BaBar Collaboration's conclusion on the $X(3915)$ quantum numbers is largely based on the assumption 
that the dominant amplitude for a $J^P=2^+$ state has helicity-2, which originally comes from quark models~\cite{Li:1990sx}. 
Abandoning this assumption the reanalysis of the data made by Zhou \emph{et al.}
concluded that the assignment $J^P=2^+$ for the $X(3915)$ is more consistent with the data, showing a sizable helicity-0 contribution in both 
$\gamma \gamma\rightarrow D\bar D$ and $\gamma \gamma\rightarrow \omega J/\psi$ amplitudes.
This large helicity contribution implies that the $X(3915)$ state might not be a pure $q\bar q$ state.
As a consequence of this analysis, PDG relabeled the resonance back to $X(3915)$, with the extra clarification: "was $\chi_{c0}(3915)$".

On the other hand, from the experimental side, a novel charmonium-like state dubbed $X(3860)$, decaying to $D\bar D$, has been reported by the Belle Collaboration~\cite{Chilikin:2017evr}, having a mass of $3862_{-32\,-13}^{+26\,+40}$ MeV/$c^2$ and a width of $201_{-67\,-82}^{+154\,+88}$ MeV. 
The $J^{PC}=0^{++}$ option is favored over the $2^{++}$ hypothesis, but its quantum numbers are not definitively determined. 
This state coincides with the suggestion of Ref~\cite{Guo:2012tv}. These authors, contrary to Belle and 
BaBar analysis, assume that all the cross section of the $\gamma\gamma\to 
D\bar{D}$ process is due to resonant structures. Therefore, the broad bump 
below the narrow peak of the $\chi_{c2}(2P)$ can be identified with the authentic 
$\chi_{c0}(2P)$, with a mass and width of $3837.6\pm11.5\,{\rm MeV}$/$c^2$ and 
$221\pm19\,{\rm MeV}$, respectively. It is worth emphasizing that the previous mass 
coincides with the predictions of some dynamical coupled-channel models~\cite{Zhou:2013ada, Danilkin:2010cc, Ortega:2012rs}.

To analyze these resonances it is necessary to take into account that, in the energy region
around 3.9 GeV/$c^2$, a significant number of open-charm channels are opened. There are convincing arguments~\cite{Pennington:2007xr,Ortega:2012rs} that open-charm thresholds 
play an important role in this energy region of the charmonium spectrum, being the charmonium-like 
resonances better described as states with a significant non-$q\bar q$ component.
Thus, the $X(3872)$ resonance together with the $X(3940)$ have been explained as two $J^{PC}=1^{++}$
states, being the $X(3872)$ basically a $D\bar D^\ast+h.c.$ molecule with a small amount of $2^3P_1$ $c\bar c$ state, while the 
$X(3940)$ is a mixture with more than $60\%$ of $c\bar c$ structure~\cite{Ortega:2010qq}. These compositenesses are essential to reproduce their properties.
Taking into account that the $X(3915)$, the $X(3930)$ and the $Y(3940)$ resonances belong to the same energy region it is reasonable to assume that the nature of 
these states are determined by the interplay between two and four quark channels.

In view of these arguments, this work explores the possible non-$q\bar q$ components of the $X(3915)$, the $X(3930)$ and the $Y(3940)$
as suggested by Zhou \emph{et al.}~\cite{Zhou:2015uva}. For that purpose we perform a coupled-channels calculation in the framework of the  
constituent quark model (CQM) proposed in Ref.~\cite{Vijande:2004he}. 
This model has been extensively used to describe the hadron phenomenology both in the light~\cite{PhysRevC.64.058201}
and the heavy quark sectors~\cite{Segovia:2016xqb}.   

The basis of the aforestated CQM is the emergence of the light-quark constituent mass as a 
consequence of the dynamical chiral symmetry breaking in QCD at some 
momentum scale. Regardless of the breaking mechanism, the simplest Lagrangian 
which describes this situation must contain Goldstone-boson fields to 
compensate the mass term. In the heavy quark sector chiral symmetry is explicitly broken and 
Goldstone-boson exchanges do not appear. However, it constrains the model 
parameters through the light-meson phenomenology~\cite{Segovia:2008zza} and 
provides a natural way to incorporate the pion exchange interaction in the 
molecular dynamics.

The potential coming from the Goldstone-boson fields is supplemented by a screened linear confinement potential
and the one-gluon exchange interaction. A scale 
dependent quark-gluon coupling constant  $\alpha_{s}$~\cite{Vijande:2004he} 
allows a consistent description of light, strange and heavy mesons (see 
Refs.~\cite{Valcarce:2005em,Segovia:2013wma} for review).
 
To find the quark-antiquark bound states  we solve the 
Schr\"odinger equation, following Ref.~\cite{Hiyama:2003cu}, we employ Gaussian trial functions with
ranges in geometric progression. This enables the optimization of ranges
employing a small number of free parameters. Moreover, the geometric
progression is dense at short distances, so that the description of
the dynamics mediated by short range potentials is properly treated. Additionally, the fast
damping Gaussian tail generated by this method can represent a problem for describing the long range. 
Fortunately, this issue can be easily overcome by choosing the maximal range much larger than the hadronic size.

In order to explore the $J^{PC}\!\!=\!0^{++}$ and $2^{++}$ charmonium sectors we employ the coupled-channels formalism described 
in Ref.~\cite{Ortega:2012rs}. We assume that the hadronic state is
\begin{equation} \label{ec:funonda}
 | \Psi \rangle = \sum_\alpha c_\alpha | \psi_\alpha \rangle
 + \sum_\beta \chi_\beta(P) |\phi_A \phi_B \beta \rangle,
\end{equation}
where $|\psi_\alpha\rangle$ are $c\bar c$ eigenstates of the two body
Hamiltonian, 
$\phi_{M}$ are $q\bar q$  eigenstates describing 
the $A$ and $B$ mesons, 
$|\phi_A \phi_B \beta \rangle$ is the two meson state with $\beta$ quantum
numbers coupled to total $J^{PC}$ quantum numbers
and $\chi_\beta(P)$ is the relative wave 
function between the two mesons in the molecule. 

In the framework of the CQM, we can derive the meson-meson potential from the $q\bar q$ interaction using the Resonating Group Method~(RGM).
For this work, the possible interactions include a direct potential, which connects open-charm meson channels,

\begin{eqnarray}
V_D&=&\sum_{i\in A;j\in B}\int \Psi^*_{l_A'm_A'}(\vec{p}_A')\Psi^*_{l_B'm_B'}(\vec{p}_B')V^D_{ij}(\vec{P}',\vec{P})\times\nonumber\\&\times&\Psi_{l_Am_A}(\vec{p}_A)\Psi_{l_Bm_B}(\vec{p}_B),
\end{eqnarray}

and an exchange one, 

\begin{eqnarray}
V_E&=&\sum_{i\in A,j\in B}\int \Psi^*_{l_A'm_A'}(\vec{p}_A')\Psi^*_{l_B'm_B'}(\vec{p}_B')V^E_{ij}(\vec{P}',\vec{P})\times\nonumber\\&\times&\Psi_{l_Am_A}(\vec{p}_A)\Psi_{l_Bm_B}(\vec{p}_B),
\end{eqnarray}
which describes the coupling between open-charm meson channels and $J/\psi\omega$, done by simple quark rearrangement driven by the $q\bar q$ interaction (see Ref.~\cite{Ortega:2010qq} for more details).

In this formalism, two- and four-quark configurations 
are coupled using the same transition mechanism that, within our approach, 
allows us to compute open-flavor meson strong decays, namely the $^{3}P_{0}$
model~\cite{LeYaouanc:1972ae, LeYaouanc:1973xz}. This model assumes that the transition operator
is 

\begin{eqnarray}
T&=&-3\sqrt{2}\gamma'\sum_\mu \int d^3 p d^3p' \,\delta^{(3)}(p+p')\times\nonumber\\&\times&
\left[ \mathcal Y_1\left(\frac{p-p'}{2}\right) b_\mu^\dagger(p)
d_\nu^\dagger(p') \right]^{C=1,I=0,S=1,J=0},
\label{TBon}
\end{eqnarray}
where $\mu$ ($\nu=\bar \mu$) are the quark (antiquark) quantum numbers and
$\gamma'=2^{5/2} \pi^{1/2}\gamma$ with $\gamma= \frac{g}{2m}$ is a dimensionless constant 
that gives the strength of 
the $q\bar q$ pair creation from the vacuum. From this operator we define the transition 
potential $h_{\beta \alpha}(P)$ within the $^{3}P_{0}$ model 
as~\cite{Kalashnikova:2005ui} 
\begin{equation}
\langle \phi_{A} \phi_{B} \beta | T | \psi_\alpha \rangle =
P \, h_{\beta \alpha}(P) \,\delta^{(3)}(\vec P_{\rm cm}).
\label{eq:Vab}
\end{equation}

The usual version of the $^{3}P_{0}$ model gives vertices that are too hard, 
specially when working at high momenta. Following the suggestion of 
Ref.~\cite{Morel:2002vk}, we use a momentum dependent form factor to truncate 
the vertex as
\begin{equation}
\label{Vab mod}
h_{\beta \alpha}(P)\to h_{\beta \alpha}(P)\times 
e^{-\frac{P^2}{2\Lambda^2}} \,,
\end{equation}
where $\Lambda=0.84\,{\rm GeV}$ is the value used herein~\cite{Ortega:2016mms}.

Using the latter coupling mechanism, the coupled-channels system can be expressed as a Schr\"odinger-type equation,

\begin{equation}
\begin{split}
\sum_{\beta} \int \big( H_{\beta'\beta}(P',P) + &
V^{\rm eff}_{\beta'\beta}(P',P) \big) \times \\
&
\times \chi_{\beta}(P) {P}^2 dP = E \chi_{\beta'}(P'),
\label{ec:Ec1}
\end{split}
\end{equation}
where $\chi_\beta(P)$ is the meson-meson relative wave function for channel $\beta$ and $H_{\beta'\beta}$ is the RGM Hamiltonian for the two-meson states obtained from
the $q\bar{q}$ interaction. The effective potential $V^{\rm eff}_{\beta'\beta}$ encodes the coupling with the $c\bar c$ bare spectrum, and can be written as

\begin{equation}
V^{\rm eff}_{\beta'\beta}(P',P;E)=\sum_{\alpha}\frac{h_{\beta'\alpha}(P')
h_{\alpha\beta}(P)}{E-M_{\alpha}},
\end {equation}
where $M_\alpha$ are the masses of the bare $c\bar{c}$ mesons.

This potential has two general effects. On the one hand, it adds additional attraction or repulsion to the $q\bar q$ interaction provided by the RGM potentials via the exchange of intermediate $c\bar c$ bare states between the two interacting mesons, which can generate new states, as it is the case for the $X(3872)$~\cite{Ortega:2010qq}. On the other hand,
the bare charmonium spectrum is renormalized by the presence of nearby meson-meson channels. 

Alternatively, Eq.~\eqref{ec:Ec1} can be solved by means of the $T$ matrix~\cite{Ortega:2012rs}, solution of the Lippmann-Schwinger equation, which is more convenient for such states above thresholds. 
Resonances will appear as poles of the $T$ matrix, namely as zeros of the inverse propagator of the mixed state, defined as
\begin{equation}
        \label{ec:pole}
        \Delta_{\alpha'\alpha}(\bar{E})=(\bar{E}-M_{\alpha})
        \delta^{\alpha'\alpha}+\mathcal{G}^{\alpha'\alpha}(\bar{E}),
\end{equation}
with $\bar{E}$ the pole position and $\mathcal{G}^{\alpha'\alpha}$ the complete mass-shift of the coupled-channels state, written as

\begin{equation} \label{ec:MshiftExact}
\begin{array}{rcl}
\mathcal{G}^{\alpha'\alpha}(E)=\sum_\beta \int dq q^2\dfrac{\phi^{\alpha\beta}(q,E)h_{\beta \alpha'}(q)}{q^2/2\mu-E},
\end{array}
\end{equation}
where $\phi^{\alpha\beta}$ are the $^3P_0$ verteces dressed by the RGM meson-meson interaction~\cite{Baru:2010ww}.

This equivalent formalism leads to a more appropriate definition of branching ratios and partial widths, following Ref.~\cite{Grassi:2000dz}. The detailed derivation has been described in Ref.~\cite{Ortega:2012rs}, so here we will only summarize the most relevant aspects. The coupled-channels $S$ matrix for an arbitrary number of $c\bar c$ states can be expressed as

\begin{eqnarray}
 S^{\beta'\beta}(E)&=&
S^{\beta'\beta}_{bg}(E)
 - \rm 2\pi\delta^{4}(P_f-P_i)\times\nonumber\\&\times&
 \sum_{\alpha,\alpha'} 
\phi^{\beta'\alpha'}(k;E)\Delta_{\alpha'\alpha}(E)^{-1}{\phi}^{\alpha\beta}(k;E),
\end{eqnarray}
where $k$ is the on-shell momentum of the two meson state and $S^{\beta'\beta}_{bg}(E)$ is the non-resonant term. Then, in the neighborhood of the pole $\bar E$, the $S$ matrix can be approximated as

\begin{eqnarray}
S^{\beta'\beta}(E)&=&
S^{\beta'\beta}_{bg}(E)
- \rm 2\pi\delta^{4}(P_f-P_i) \times\nonumber\\&\times&\sum_{\alpha,\alpha'} 
\phi^{\beta'\alpha'}(\bar k;\bar E)\frac{\mathcal Z_{\alpha'\alpha}(\bar E)^{-1}}{E-\bar E}
{\phi}^{\alpha\beta}(\bar k;\bar E),
\end{eqnarray}
where
\begin{equation}
 Z^{\alpha'\alpha}(\bar E)= \lim_{E\to \bar E} \frac{\Delta^{\alpha'\alpha}(E)-\Delta^{\alpha'\alpha}(\bar E)}{E-\bar E}.
 \end{equation}

So, assuming that we can write $\mathcal Z_{\alpha'\alpha}(\bar E)=\sum_{\lambda}\mathcal Z_{\alpha'\lambda}^{1/2}\mathcal Z_{\lambda\alpha}^{1/2}$ the $S$ matrix is, finally
\begin{eqnarray}
S^{\beta'\beta}(E)&=&
S^{\beta'\beta}_{bg}(E)
- \rm 2\pi\delta^{4}(P_f-P_i) \nonumber\times\\&\times&\sum_{\alpha,\alpha',\lambda} 
\left[\phi^{\beta'\alpha'}(\bar k;\bar E)\mathcal 
Z_{\alpha'\lambda}(E)^{-1/2}\right]\frac{1}{E-\bar E}\times \nonumber\\
&\times&\left[\mathcal 
Z_{\lambda\alpha}(E)^{-1/2}{\phi}^{\alpha\beta}(\bar k;\bar E)\right],
\end{eqnarray}
where we can identify the decay vertex
\begin{equation}
S(X_c\rightarrow f)^{\beta\alpha}= \sum_{\lambda}\phi^{\beta\lambda}(\bar 
k;\bar E)\mathcal Z_{\lambda\alpha}(\bar E)^{-1/2}.
\end{equation}

From there, the partial width of a two meson decay $\hat \Gamma_\beta$ can be written as

\begin{eqnarray}
\label{ec:parcial2}
 \hat\Gamma_\beta&=&2\pi \frac{E_1E_2}{M_r}{k}_\beta
\sum_{\alpha',\alpha,\lambda}{\phi^*}^{\beta\alpha'}(\bar k)\mathcal 
Z^*_{\alpha'\lambda}(\bar E)^{-1/2} \times\nonumber\\&\times&\mathcal Z(\bar E)^{-1/2}_{\lambda 
\alpha}\phi^{\alpha\beta}(\bar k),
\end{eqnarray}
where $\bar E=M_r-i\tfrac{\Gamma_r}{2}$, $k_\beta$ is the on-shell momentum for the meson-meson $\beta$ channel and
$E_i$ is the on-shell total energy of mesons $i=\{1,2\}$.

The previous equation~(\ref{ec:parcial2}) does not, in general, satisfy that the sum of the partial 
widths must be equal to the total width. This issue can be easily solved by defining the 
branching ratios as~\cite{Grassi:2000dz}
\begin{equation} \label{ec:bratios}
\mathcal{B}_f=\frac{\hat\Gamma_f}{\sum_f\hat \Gamma_f},
\end{equation}
so the physical partial widths are $\Gamma_f=\mathcal{B}_f \Gamma_r$.

We have performed two calculations for the quantum numbers $J^{PC}\!\!=\!0^{++}$ and $J^{PC}\!\!=\!2^{++}$. The first one includes, for the $J^{PC}\!\!=\!0^{++}$ charmonium sector,
the naive $2^3P_0$ $c\bar c$ state together with the following channels (their corresponding threshold energies are indicated in parenthesis): $D\bar D$ ($3734$ MeV/$c^2$), $\omega J/\psi$ ($3880$ MeV/$c^2$), $D_s\bar D_s$ ($3937$ MeV/$c^2$) and $D^\ast\bar D^\ast$ ($4017$ MeV/$c^2$). 
For the $J^{PC}=2^{++}$ case we add to the former channels the $D\bar D^\ast+h.c.$ ($3877$ MeV/$c^2$) one, which in this case will be coupled to the bare $2^3P_2$ $c\bar c$ state. These thresholds have been considered 
because of their closeness to the masses of the naive $2^3P_J$ (J=0,2) states predicted by the quark model. Moreover, the $D^\ast\bar D^\ast$ threshold, though located at higher energies compared to the other channels, must be included because it is the only one contributing with an $S-$wave in the $J^{PC}\!\!=\!0^{++}$ sector and can have a major impact on the dynamics of the system. Its inclusion for the $J^{PC}\!\!=\!2^{++}$ case is needed to compare both sectors.

Using the original parameters of Ref.~\cite{Ortega:2010qq} (which will be denoted as {\it model A}) we obtain the masses and widths shown in Table~\ref{tab:model A}.

\begin{table*}[!t]
\caption{\label{tab:model A} Mass and decay width, in MeV, and probabilities of 
the different Fock components, for model A.}
\begin{ruledtabular}
\begin{tabular}{ccccccccc}
$J^{PC}$ & Mass & Width & ${\cal P} [c\bar{c}] $ & ${\cal P}[D\bar D]$ & ${\cal P}[D \bar D^{\ast}]$ & ${\cal 
P}[\omega J/\psi]$ & ${\cal P}[D_s\bar D_s]$ & ${\cal P}[D^{\ast} \bar D^{\ast}]$\\
\hline
$0^{++}$ & $3890.3$ & $6.7$ & $44.1\%$ & $21.6\%$ & $-$ & $28.4\%$ & $2.6\%$ & $3.3\%$ \\
$0^{++}$ & $3927.4$ & $229.8$ & $19.2\%$ & $66.3\%$ & $-$ & $5.3\%$ & $3.7\%$ & $5.5\%$ \\
$2^{++}$ & $3925.6$ & $19.0$ & $42.2\%$ & $11.3\%$ & $37.0\%$ & $4.0\%$ & $0.4\%$ & $5.1\%$ \\

\end{tabular}
\end{ruledtabular}
\end{table*}

We find two states with $J^{PC}\!\!=\!0^{++}$ and only one with $J^{PC}\!\!=\!2^{++}$ because the interaction in the meson-meson channel for the latter sector is not strong enough
to generate a second resonance. The mass and width of the $J^{PC}\!\!=\!2^{++}$ state is compatible with those of the $X(3930)$, whereas the mass 
of the first $J^{PC}\!\!=\!0^{++}$ state is more similar to the new $X(3860)$ resonance than the one of the $X(3915)$. However, our width is smaller than the experimental one. 
Such small value is connected with the position of the node in the $2^3P_0$ bare wave function, which affects the $^3P_0$ transition 
amplitudes and, hence, causes a higher sensitivity of the width to small changes in the wave function structure or, alternatively, the mass of the $X(3860)$ resonance. 
A recent analysis of the decay width of the X(3860) has been performed by Ref.~\cite{Yu:2017bsj}, using a simple harmonic oscillator (SHO) approximation for the meson wave function. The X(3860) width shows a strong dependence with the oscillator parameter, finding agreement with the experimental data with a resonable value.
In our case, all the parameters are fixed by the strong decays of light and heavy quark mesons~\cite{Segovia:2012cd} and the $q\bar q$ dynamics and, thus, a similar fine-tuning cannot be done.

The mass of the second $J^{PC}\!\!=\!0^{++}$ state allows us to assign it to the $Y(3940)$ resonance. However, as in the former case, its width is far from the experimental value. 
This disagreement in the width of both states suggests a new, that may be more interesting, assignment. One can identify the second $0^{++}$ state with the $X(3860)$, as the width of the state ($229.8$ MeV in Table~\ref{tab:model A}) matches with the experimental data, whereas, considering that the measured mass even reaches more than $3900$ MeV, the discrepancy of the experimental mass value with the theoretical one is within the range of the uncertainties of the model. Additionally, the extra state with a width of $6.7$ MeV is too narrow and can hardly be observed in the experiment of Ref.~\cite{Chilikin:2017evr}.  With the assignment of the $X(3860)$ to the broader $0^{++}$ resonance, we do not find any candidate to the $Y(3940)$ signal, which would be in agreement with BaBar suggestion that this resonance is the same as the $X(3915)$~\cite{PhysRevLett.101.082001}.

 Certainly, all the states show a sizable no-$q\bar q$ structure and therefore cannot be assigned to pure $q\bar q$ states. This fact overrides the concern about the hyperfine splitting because the masses of the $q\bar q$ states are renormalized by the coupling with the different meson-meson channels. 

To explore the robustness of the results, taken into account the uncertainties of the model parameters, we have performed a second calculation (named {\it model B}) where we have slightly changed the bare mass of 
the $2^3P_J$ $c\bar c$ pairs ($0.25\%$) and used the coupling of the $^3P_0$ model from Ref.~\cite{Segovia:2012cd}, which represent a change from $\gamma= 0.226$ to $\gamma=0.286$ for the charmonium sector. The results of the new calculation are shown in Table~\ref{tab:model B}.

\begin{table*}[!t]
\caption{\label{tab:model B} Mass and decay width, in MeV, and probabilities of 
the different Fock components for model B.}
\begin{ruledtabular}
\begin{tabular}{ccccccccc}
$J^{PC}$ & Mass & Width & ${\cal P} [c\bar{c}]$ & ${\cal P}[D\bar D]$ & ${\cal P}[D \bar D^{\ast}]$ & ${\cal 
P}[\omega J/\psi]$ & ${\cal P}[D_s\bar D_s]$ & ${\cal P}[D^{\ast} \bar D^{\ast}]$\\
\hline
$0^{++}$ & $3889.0$ & $11.8$ & $43.5\%$ & $27.3\%$ & $-$ & $20.4\%$ & $3.8\%$ & $4.9\%$ \\
$0^{++}$ & $3947.5$ & $201.6$ & $19.4\%$ & $66.0\%$ & $-$ & $3.7\%$ & $8.0\%$ & $2.9\%$ \\
$2^{++}$ & $3915.1$ & $19.8$ & $37.8\%$ & $14.1\%$ & $36.4\%$ & $5.12\%$ & $0.4\%$ & $6.1\%$ \\

\end{tabular}
\end{ruledtabular}
\end{table*}

Interestingly, this new parametrization leads to practically the same results for the first $J^{PC}\!\!=\!0^{++}$ state and the same compositeness for the
$J^{PC}\!\!=\!2^{++}$, although now the mass is more similar to the $X(3915)$ resonance. The mass of the second $J^{PC}\!\!=\!0^{++}$ state is slightly increased, 
although such modification is of the order of the experimental error of the $Y(3940)$ resonance.

In view of these results, we can proceed and calculate for the $J^{PC}=2^{++}$ state the product of the two-photon decay width and the branching fraction to $\omega J/\psi$ and 
$D\bar D$ channels, assuming the $X(3915)$ and $X(3930)$ are the same $J^{PC}\!\!=\!2^{++}$ resonance. 
The results are quoted in Table~\ref{tab:BR J2} where we also include the decay to the $D\bar D^*$ channel.

Our model predicts a value for the branching fraction of the $2^{++}$ state to $D\bar D$ some standard deviations below the experimental one. 
This value is obtained from the decay to 
the $I=0$ $D\bar D$ channel as incorporated in the coupled-channels calculation. However, it does not include possible contributions from higher open-charm
channels decaying to $D\bar D$ pairs, such as the decay of $D^\ast$ to $D\gamma$ or $D\pi$ in the $D\bar D^\ast$ channel. 
As shown in Table~\ref{tab:BR J2}, our calculated value for the branching fraction to $D\bar D^\ast$ channel is higher than the one for $D\bar D$, so it is reasonable
to assume that part of the $D\bar D^\ast$ pairs decaying to $D\bar D\gamma$ and $D\bar D\pi$ are, in fact, measured as $D\bar D$ pairs, increasing 
our theoretical branching fraction for the $D\bar D$ channel. Under this assumption, the disagreement between our value and the experimental branching fraction
can be easily explained if just one third of the $D\bar D^\ast$ decays are measured as $D\bar D$ pairs.

\begin{table*}[!t]
\caption{\label{tab:BR J2} Product of the two-photon decay width and the branching fraction to different channels (in eV) for the 
$J^{PC}=2^{++}$ sector for each model, and comparison with Belle and BaBar Collaboration experimental results.}
\begin{ruledtabular}
\begin{tabular}{ccccc}
 & Belle & BaBar & model A &  model B  \\
\hline
$\Gamma_{\gamma \gamma} \times {\cal B}(2^{++}\to \omega J/\psi )$ & $18\pm 5 \pm 2$~\cite{Uehara:2009tx} & $10.5\pm1.9\pm 0.6$~\cite{Lees:2012xs} & $20.9$ & $24.9$  \\
$\Gamma_{\gamma \gamma} \times {\cal B}(2^{++}\to D\bar D )$ & $180\pm 50\pm 30 $~\cite{Uehara:2005qd} & $249\pm 50\pm 40$~\cite{Aubert:2010ab} & $75.4$ & $81.4$  \\
$\Gamma_{\gamma \gamma} \times {\cal B}(2^{++}\to D\bar{D^\ast})$ & - & - & $196.0$ & $151.9$  \\

\end{tabular}
\end{ruledtabular}
\end{table*}

As indicated by Table~\ref{tab:BR J2}, the results for both model A and B are very similar and not far from the experimental data. Then, both models describe the experimental branchings providing that the $X(3915)/X(3930)$ resonances are $J^{PC}\!\!=\!2^{++}$. This conclusion agrees with Ref.~\cite{Branz:2010rj}. 

Assuming the assignment of the broader resonance to the $Y(3940)$, we can estimate the product branching function $\mathcal B(B\to KY(3940))\times\mathcal B(Y(3940)\to \omega J/\psi)$. Following Olsen~\cite{Olsen:2014maa}, we can assume that, due to the significant $\chi_{c0}(2P)$ component,  the $\mathcal B(B\to KY(3940))$ should be less than or equal to $\mathcal B(B\to K\chi_{c0}(1P))$. This assumption is based on the fact that the width of P-wave mesons is proportional to the derivative  of the $q\bar q$ radial wave function at the origin, which decreases with increasing radial excitation. Moreover, the available phase space is smaller. With this assumption we obtain 
 $\mathcal B(B\to KY(3940))\times\mathcal B(Y(3940)\to \omega J/\psi)\leq3.3\times 10^{-5}$ for the model A and $\mathcal B(B\to KY(3940))\times\mathcal B(Y(3940)\to \omega J/\psi)\leq 2.9\times 10^{-5}$ for the model B, which in both cases is of the same order of magnitude as the experimental result, $(7.1\pm 1.3\pm 3.1)\times 10^{-5}$~\cite{PhysRevLett.94.182002}.

In summary, within a coupled-channels calculation we have obtained two $J^{PC}\!\!=\!0^{++}$ and one $J^{PC}\!\!=\!2^{++}$ resonances in the energy region of $3.9$ GeV/$c^2$. Using the parametrization of Ref~\cite{Ortega:2012rs} we obtain two possible description of the charmonium-like states experimentally measured in this region. On the one hand, the $X(3860)$ is identified with the second $J^{PC}\!\!=\!0^{++}$ state, with the right width but slightly higher mass, % the two $J^{PC}\!\!=\!1^{++}$ resonances with the $X(3872)$ and the $X(3940)$ 
and the $J^{PC}\!\!=\!2^{++}$ state with the $X(3915)/X(3930)$. On the second hand, the two $J^{PC}\!\!=\!0^{++}$ states  are identify with the 
$X(3860)$ and the $Y(3940)$, maintaining the assignment for the other resonances. Including the results of Ref.~\cite{Ortega:2012rs} for the $J^{PC}\!\!=\!1^{++}$ charmonium sector, where two resonances, the $X(3872)$ and the $X(3940)$, are described, the present work completes the picture of the P-wave charmonia around $3.9$ GeV/$c^2$. All these  states are mixtures of $\chi_{cJ}(2P)$ charmonium states and meson-meson channels. Therefore neither can be identified with pure $c\bar c$ states, which explains their deviations from the naive quark model predictions. Among other characteristics, this compositeness is able to explain the properties of the $X(3872)$~\cite{Ortega:2012rs}. 

Within the uncertainties of our model, the mass and width of the $J^{PC}\!\!=\!2^{++}$ state can be identified either with the $X(3930)$ or with the $X(3915)$, suggesting that the two resonances $X(3915)$ and $X(3930)$ are in fact the same $J^{PC}\!\!=\!2^{++}$ as claimed by Z.-Y. Zhou \emph{et al.}~\cite{Zhou:2015uva}. 
We may identify  the new $X(3860)$ resonance with a $J^{PC}\!\!=\!0^{++}$ as suggested in Ref~\cite{Guo:2012tv}. Finally, in the second scenario we find a resonance which reproduces the experimental data of the  $Y(3940)$ as a $J^{PC}\!\!=\!0^{++}$, which may encourage new experimental searches for this state.
In any case, further theoretical and experimental work is necessary to fully unveil the nature of these $c\bar c$ resonances in this energy region.

%%%%%%%%%%%%%%%%%%%%%%%%%%%%%%%%%%%%%%%%%%%%%%%%%%%%%%%%%%%%%%%%%%%%%%%%%%%%%%%

\begin{acknowledgments}
This work has been partially funded by Ministerio de Ciencia y Tecnolog\'ia 
under Contract no. FPA2013-47443-C2-2-P, by Ministerio de Econom\'ia, Industria y 
Competitividad under
Contract no. FPA2016-77177-C2-2-P, and by Junta de Castilla y Le\'on 
and European Regional Development Funds (ERDF) under Contract no. SA041U16. 
J.S. acknowledges the financial support from Alexander von Humboldt Foundation.
\end{acknowledgments}

%%%%%%%%%%%%%%%%%%%%%%%%%%%%%%%%%%%%%%%%%%%%%%%%%%%%%%%%%%%%%%%%%%%%%%%%%%%%%%%
%%%%%%%%%%%%%%%%%%%%%%%%%%%%%%%%%%%%%%%%%%%%%%%%%%%%%%%%%%%%%%%%%%%%%%%%%%%%%%%

\bibliography{paperX3915}

\end{document}